# Nonlinear field theory II

## The field representation of the Dirac electron theory

Alexander G. Kyriakos

Saint-Petersburg State Institute of Technology,
St. Petersburg, Russia*)

**Abstract**

The present paper is the continuation of the paper "Nonlinear field theory I". In the paper it is shown that a fully correspondence between the quantum and the electromagnetic forms of the Dirac electron theory exists, so that each element of the Dirac theory acquires the known electrodynamics meaning and reversely.

**PACS** numbers:  12.20.–m  Quantum electrodynamics

## 1.0. Introduction

On the basis of the previous paper "The non-linear field theory I" [A. G. Kyriakos. http://arXiv.org/abs/quant-ph/0404088] we will show here that all the mathematical particularities of the Dirac electron theory have the known electrodynamics meaning.

## 2.0. Electrodynamics meaning of the forms of the Dirac equations

There are two bispinor Dirac equations [1,2,3]:

$$[(\hat{\alpha}_o \hat{\varepsilon} + c\hat{\vec{\alpha}} \, \hat{\vec{p}}) + \hat{\beta} \, mc^2]\psi = 0, \qquad (1.1)$$

$$\psi^+[(\hat{\alpha}_o \hat{\varepsilon} - c\hat{\vec{\alpha}} \, \hat{\vec{p}}) - \hat{\beta} \, mc^2] = 0, \qquad (1.2)$$

which correspond to two signs of the relativistic expression of the electron energy:

$$\varepsilon = \pm\sqrt{c^2 \vec{p}^2 + m^2 c^4}, \qquad (1.3)$$

Here $\hat{\varepsilon} = i\hbar \dfrac{\partial}{\partial t}$, $\hat{\vec{p}} = -i\hbar \vec{\nabla}$ are the operators of the energy and momentum, $\varepsilon, \vec{p}$ are the electron energy and momentum, $c$ is the light velocity, $m$ is the electron mass, $\psi^+$ is the Hermitian-conjugate wave function and $\hat{\alpha}_o = \hat{1}$, $\hat{\vec{\alpha}}$, $\hat{\alpha}_4 \equiv \hat{\beta}$ are the Dirac matrices:

$$\hat{\alpha}_0 = \begin{pmatrix} \hat{\sigma}_0 & 0 \\ 0 & \hat{\sigma}_0 \end{pmatrix}, \; \hat{\vec{\alpha}} = \begin{pmatrix} 0 & \hat{\vec{\sigma}} \\ \hat{\vec{\sigma}} & 0 \end{pmatrix}, \; \hat{\beta} \equiv \hat{\alpha}_4 = \begin{pmatrix} \hat{\sigma}_0 & 0 \\ 0 & -\hat{\sigma}_0 \end{pmatrix}, \qquad (1.4)$$

where $\hat{\vec{\sigma}}$ are Pauli spin matrices: $\hat{\sigma}_1 = \begin{pmatrix} 0 & 1 \\ 1 & 0 \end{pmatrix}$, $\hat{\sigma}_2 = \begin{pmatrix} 0 & -i \\ i & 0 \end{pmatrix}$, $\hat{\sigma}_3 = \begin{pmatrix} 1 & 0 \\ 0 & -1 \end{pmatrix}$, $\hat{\sigma}_0 = \begin{pmatrix} 1 & 0 \\ 0 & 1 \end{pmatrix}$.

(Note that for each sign of the equation (1.3) there are two Hermitian-conjugate Dirac equations).

### 2.1. The Dirac equation without mass

Let us consider the plane electromagnetic wave moving, for example, on $y$ - axis in the complex form:

$$\begin{cases} \vec{E} = \vec{E}_o e^{-i(\omega t \pm ky)}, \\ \vec{H} = \vec{H}_o e^{-i(\omega t \pm ky)}, \end{cases} \qquad (2.1)$$

We will suppose that the complex form of electromagnetic waves corresponds to the complexity of the Dirac wave functions. In the general case the electromagnetic wave, moving on $y$ - axis, has two polarizations and contains the following four field vectors:

$$(E_x, E_z, H_x, H_z) \qquad (2.2)$$

As in this case $E_y = H_y = 0$ for all transformations, we can compare the set (2.2) with the Dirac wave function. Let's now consider the electromagnetic wave equation:

$$\left(\dfrac{\partial^2}{\partial t^2} - c^2 \vec{\nabla}^2\right)\vec{F} = 0, \qquad (2.5)$$

where $\vec{F}$ is whichever of the electromagnetic wave functions, particularly, the fields (2.2). In other words this equation represents four equations for each wave function of the electromagnetic field.

In case of the wave, moving along the $y$ - axis, we can write this equation in the following form:

$$\left(\hat{\varepsilon}^2 - c^2 \hat{\vec{p}}^2\right) \vec{F}(y) = 0, \tag{2.6}$$

The equation (2.6) can also be represented in the form of the Klein-Gordon equation[6] without mass:

$$\left[\left(\hat{\alpha}_o \hat{\varepsilon}\right)^2 - c^2 \left(\hat{\vec{\alpha}} \ \hat{\vec{p}}\right)^2\right] \psi = 0, \tag{2.7}$$

where $\psi$ is a matrix, which in some way consists of the fields (2.2). In fact, taking into account that

$$\left(\hat{\alpha}_o \hat{\varepsilon}\right)^2 = \hat{\varepsilon}^2, \quad \left(\hat{\vec{\alpha}} \ \hat{\vec{p}}\right)^2 = \hat{\vec{p}}^2, \tag{2.8}$$

we can realize that the equations (2.6) and (2.7) are equivalent.

Factorizing (2.7) and multiplying it from the left on the Hermitian-conjugate function $\psi^+$ we get:

$$\psi^+ \left(\hat{\alpha}_o \hat{\varepsilon} - c\hat{\vec{\alpha}} \ \hat{\vec{p}}\right) \left(\hat{\alpha}_o \hat{\varepsilon} + c\hat{\vec{\alpha}} \ \hat{\vec{p}}\right) \psi = 0, \tag{2.9}$$

The equation (2.9) may be disintegrated on two Dirac equations without mass:

$$\psi^+ \left(\hat{\alpha}_o \hat{\varepsilon} - c\hat{\vec{\alpha}} \ \hat{\vec{p}}\right) = 0, \tag{2.10}$$

$$\left(\hat{\alpha}_o \hat{\varepsilon} + c\hat{\vec{\alpha}} \ \hat{\vec{p}}\right) \psi = 0, \tag{2.11}$$

It is not difficult to show that only in case we choose the Dirac bispinors as

$$\psi = \begin{pmatrix} E_x \\ E_z \\ iH_x \\ iH_z \end{pmatrix}, \quad \psi^+ = \begin{pmatrix} E_x & E_z & -iH_x & -iH_z \end{pmatrix}, \tag{2.12}$$

the *equations (2.10) and (2.11) are the correct Maxwell equations of the electromagnetic waves: retarded and advanced.*

## 2.2. The Dirac equation form with mass

The full Dirac equation has a lot of equivalent forms. Let us consider first two Hermitian-conjugate equations, corresponding to the minus sign of the expression (1.9):

$$\left[\left(\hat{\alpha}_o \hat{\varepsilon} + c\hat{\vec{\alpha}} \ \hat{\vec{p}}\right) + \hat{\beta} \ mc^2\right] \psi = 0, \tag{2.13'}$$

$$\psi^+ \left[\left(\hat{\alpha}_o \hat{\varepsilon} + c\hat{\vec{\alpha}} \ \hat{\vec{p}}\right) + \hat{\beta} \ mc^2\right] = 0, \tag{2.13''}$$

Using (2.12), from (2.13') and (2.13'') we obtain:

$$\begin{cases} \dfrac{1}{c}\dfrac{\partial E_x}{\partial t} - \dfrac{\partial H_z}{\partial y} = -\vec{j}^{\,e}_x \\ \dfrac{1}{c}\dfrac{\partial H_z}{\partial t} - \dfrac{\partial E_x}{\partial y} = \vec{j}^{\,m}_z \\ \dfrac{1}{c}\dfrac{\partial E_z}{\partial t} + \dfrac{\partial H_x}{\partial y} = -\vec{j}^{\,e}_z \\ \dfrac{1}{c}\dfrac{\partial H_x}{\partial t} + \dfrac{\partial E_z}{\partial y} = \vec{j}^{\,m}_x \end{cases} \quad (2.14'), \qquad \begin{cases} \dfrac{1}{c}\dfrac{\partial E_x}{\partial t} - \dfrac{\partial H_z}{\partial y} = \vec{j}^{\,e}_x \\ \dfrac{1}{c}\dfrac{\partial H_z}{\partial t} - \dfrac{\partial E_x}{\partial y} = -\vec{j}^{\,m}_z \\ \dfrac{1}{c}\dfrac{\partial E_z}{\partial t} + \dfrac{\partial H_x}{\partial y} = \vec{j}^{\,e}_z \\ \dfrac{1}{c}\dfrac{\partial H_x}{\partial t} + \dfrac{\partial E_z}{\partial y} = -\vec{j}^{\,m}_x \end{cases} \quad (2.14'')$$

where
$$\vec{j}^{\,e} = i\dfrac{\omega}{4\pi}\vec{E} = i\dfrac{1}{4\pi}\dfrac{c}{r_C}\vec{E}, \qquad (2.15')$$

$$\vec{j}^{\,m} = i\dfrac{\omega}{4\pi}\vec{H} = i\dfrac{1}{4\pi}\dfrac{c}{r_C}\vec{H}, \qquad (2.15'')$$

are the complex currents, in which $\omega = \dfrac{mc^2}{\hbar}$, and $r_C = \dfrac{\hbar}{mc}$ is the Compton length wave of the electron. Thus, the equations (2.13') and (2.13'') are Maxwell equations with complex currents. As we see, *the Hermitian-conjugate equations (2.14) and (2.15) differ by the current directions*.

It is interesting that together with the electric current $\vec{j}^{\,e}$ the magnetic current $\vec{j}^{\,m}$ also exists here. The last one must be equal to zero according to the Maxwell theory, but its existence according to Dirac doesn't contradict to the quantum theory (see the Dirac theory of the magnetic monopole).

Let us consider now the equations that correspond to minus signs of (1.9):

$$\left[ \left( \hat{\alpha}_o \hat{\varepsilon} - c\hat{\vec{\alpha}}\ \hat{\vec{p}} \right) - \hat{\beta}\, mc^2 \right] \psi = 0, \qquad (2.16)$$

$$\psi^+ \left[ \left( \hat{\alpha}_o \hat{\varepsilon} - c\hat{\vec{\alpha}}\ \hat{\vec{p}} \right) - \hat{\beta}\, mc^2 \right] = 0, \qquad (2.17)$$

The electromagnetic form of the equation (2.16) is:

$$\begin{cases} \dfrac{1}{c}\dfrac{\partial E_x}{\partial t} + \dfrac{\partial H_z}{\partial y} = -\vec{j}^{\,e}_x \\ \dfrac{1}{c}\dfrac{\partial H_z}{\partial t} + \dfrac{\partial E_x}{\partial y} = \vec{j}^{\,m}_z \\ \dfrac{1}{c}\dfrac{\partial E_z}{\partial t} - \dfrac{\partial H_x}{\partial y} = -\vec{j}^{\,e}_z \\ \dfrac{1}{c}\dfrac{\partial H_x}{\partial t} - \dfrac{\partial E_z}{\partial y} = \vec{j}^{\,m}_x \end{cases}, \qquad (2.18)$$

Obviously, the electromagnetic form of the equation (2.17) will have the opposite signs of the currents comparatively to (2.18).

Comparing (2.18) and (2.14) we can see that the *equation (2.18) can be considered as the Maxwell equation of the retarded wave*. If we don't want to use the retarded wave, we can transform the wave function of the retarded wave to the form:

$$\psi_{ret} = \begin{pmatrix} E_x \\ -E_z \\ iH_x \\ -iH_z \end{pmatrix}, \qquad (2.19)$$

Then, contrary to the system (2.18) we get the system (2.15). The transformation of the function $\psi_{ret}$ to the function $\psi_{adv}$ is called the charge conjugation operation.

The Dirac electron theory has a lot of particularities. In the modern interpretation these particularities are considered as mathematical features that do not have a physical meaning. Below we will show that in the electromagnetic form all the Dirac electron theory particularities have the known physical sense.

## 3.0. Electrodynamics sense of the bispinor form of the Dirac electron theory

It is known that there are 16 Dirac matrices of 4x4 dimensions. We use the set of matrices which used Dirac himself and will name it $\alpha$-set (1.4).

It can be shown that the tensor dimension of bilinear form follows from its nonlinear electrodynamics forms. Enumerate correspondingly Dirac's matrices [1,2,3]:

1) $\hat{\alpha}_4 \equiv \hat{\beta}$

2) $\hat{\alpha}_\mu = \{\hat{\alpha}_0, \hat{\vec{\alpha}}\} \equiv \{\hat{\alpha}_0, \hat{\alpha}_1, \hat{\alpha}_2, \hat{\alpha}_3, \hat{\alpha}_4\}$

3) $\hat{\alpha}_5 = \hat{\alpha}_1 \cdot \hat{\alpha}_2 \cdot \hat{\alpha}_3 \cdot \hat{\alpha}_4$ (3.1)

4) $\hat{\alpha}_\mu^A = \hat{\alpha}_5 \cdot \hat{\alpha}_\mu$

5) $\hat{\alpha}_{\mu\nu} = -\hat{\alpha}_{\nu\mu} = \begin{cases} i\hat{\alpha}_\nu \hat{\beta} \hat{\alpha}_\mu, & \mu \neq \nu \\ 0, & \mu = \nu \end{cases}$

where 1) scalar, 2) 4-vector, 3) pseudoscalar, 4) 4-pseudovector, 5) antisymmetrical tensor of second rank are.

Let's calculate electrodynamics values corresponding to these matrices: $O = \psi^+ \hat{\alpha} \psi$, where $\psi$ is given by (2.12):

1) $\psi^+ \hat{\alpha}_4 \psi = (E_x^2 + E_z^2) - (H_x^2 + H_z^2) = \vec{E}^2 - \vec{H}^2 = 8\pi I_1$, where $I_1$ is the *first scalar* (invariant) of Maxwell theory, i.e. the Lagrangian of electromagnetic field in vacuum;

2) $\psi^+ \hat{\alpha}_o \psi = \vec{E}^2 + \vec{H}^2 = 8\pi U$, where $U$ is the *energy density* of electromagnetic field;

$\psi^+ \hat{\alpha}_y \psi = -\dfrac{8\pi}{c} S_{Py} = -8\pi\, c\vec{g}_y$, where $\vec{g}_y$ is the *momentum density* of the electromagnetic wave field moved along the Y-axis. As it is known, the value $\left\{\dfrac{1}{c} U, \vec{g}\right\}$ is 4-vector of energy-momentum.

3) $\psi^+\hat{\alpha}_5\psi = 2(E_xH_x + E_zH_z) = 2(\vec{E}\cdot\vec{H})$ which is the *pseudoscalar* of electromagnetic field, and $(\vec{E}\cdot\vec{H})^2 = I_2$ is the second scalar (invariant) of electromagnetic field theory.

4) $\psi^+\hat{\alpha}_5\hat{\alpha}_0\psi = 2(E_xH_x + E_zH_z) = 2(\vec{E}\cdot\vec{H})$

$\psi^+\hat{\alpha}_5\hat{\alpha}_1\psi = -2i(E_xE_z - H_xH_z)$,

$\psi^+\hat{\alpha}_5\hat{\alpha}_2\psi = 0$,

$\psi^+\hat{\alpha}_5\hat{\alpha}_3\psi = -i(E_x^2 - E_z^2 - H_x^2 + H_z^2)$.

5) Tensor $\psi^+\hat{\alpha}_{\mu\nu}\psi$ we can write in compact form:

$$(\alpha_{\mu\nu}) = \begin{pmatrix} 0 & E_x^2 - E_z^2 + H_x^2 - H_z^2 & 0 & -2(E_xH_z + E_zH_x) \\ -(E_x^2 - E_z^2 - H_x^2 + H_z^2) & 0 & 2(E_xE_z - H_xH_z) & 0 \\ 0 & -2(E_xE_z - H_xH_z) & 0 & -2(E_xH_x - E_zH_z) \\ 2(E_xH_z + E_zH_x) & 0 & 2(E_xH_x - E_zH_z) & 0 \end{pmatrix}$$

## 4.0. About statistical interpretation of the quantum mechanics

As it is known, from the Dirac equation the probability continuity equation can be obtained []:

$$\frac{\partial P_{pr}(\vec{r},t)}{\partial t} + div\ \vec{S}_{pr}(\vec{r},t) = 0, \qquad (4.1)$$

Here $P_{pr}(\vec{r},t) = \psi^+\hat{\alpha}_0\psi$ is the probability density, and $\vec{S}_{pr}(\vec{r},t) = -c\psi^+\hat{\vec{\alpha}}\psi$ is the probability flux density. Using the above results we can obtain: $P_{pr}(\vec{r},t) = 8\pi\ U$ and $\vec{S}_{pr} = c^2\vec{g} = 8\pi\ \vec{S}$. Then the electromagnetic form of the equation (3.15) is:

$$\frac{\partial U}{\partial t} + div\ \vec{S} = 0, \qquad (4.2)$$

which is the form of *energy conservation law* of the electromagnetic field.

## 5.0. The electrodynamics sense of the matrices choice

According to [4] "it can prove that all the physical consequences of Dirac's equation do not depend on the special choice of Dirac's matrices… Their differences are only apparent". The matrix sequence $(\hat{\alpha}_1, \hat{\alpha}_2, \hat{\alpha}_3)$ agrees with the electromagnetic wave, which has $-y$-direction. A question arises: how to describe the waves, which have $x$ and $z$ - directions? Introducing the axes' indexes, which indicate the electromagnetic wave direction, we can write three groups of the matrices, each of which corresponds to one and only one wave direction:

$(\hat{\alpha}_{1x}, \hat{\alpha}_{2y}, \hat{\alpha}_{3z})$, $(\hat{\alpha}_{2x}, \hat{\alpha}_{3y}, \hat{\alpha}_{1z},)$, $(\hat{\alpha}_{2z}, \hat{\alpha}_{1y}, \hat{\alpha}_{3x})$.

Let us choose now the wave function forms, which give the correct Maxwell equations for the $x$ and $z$ - directions. Taking into account (2.12) as the initial form of the $-y$ - direction, from it, by means of the indexes' transposition around the circle, we will get other forms. Since in this case the Poynting vector has the minus sign, we can suppose that the transposition must be counterclockwise. Let us examine this supposition, checking the Poynting vector values:
the $(\hat{\alpha}_{1x}, \hat{\alpha}_{2y}, \hat{\alpha}_{3z})$, $(\hat{\alpha}_{2x}, \hat{\alpha}_{3y}, \hat{\alpha}_{1z},)$, $(\hat{\alpha}_{2z}, \hat{\alpha}_{1y}, \hat{\alpha}_{3x})$ correspond to the wave functions

$$\psi(y) = \begin{pmatrix} E_x \\ E_z \\ iH_x \\ iH_z \end{pmatrix}, \quad \psi(x) = \begin{pmatrix} E_z \\ E_y \\ iH_z \\ iH_y \end{pmatrix}, \quad \psi(z) = \begin{pmatrix} E_y \\ E_x \\ iH_y \\ iH_x \end{pmatrix} \text{ and to non-zero Poynting vectors}$$

$\psi^+ \hat{\alpha}_{2y} \psi = -2[\vec{E} \times \vec{H}]_y$, $\psi^+ \hat{\alpha}_{2x} \psi = -2[\vec{E} \times \vec{H}]_x$, $\psi^+ \hat{\alpha}_{2z} \psi = -2[\vec{E} \times \vec{H}]_z$ respectively.

As we see, we took the correct result. We can suppose now that, by the clockwise indexes' transposition, the wave functions will describe the electromagnetic waves, which move in a positive direction along the co-ordinate axes. Let us prove this:
the $(\hat{\alpha}_{1x}, \hat{\alpha}_{2y}, \hat{\alpha}_{3z})$, $(\hat{\alpha}_{2x}, \hat{\alpha}_{3y}, \hat{\alpha}_{1z})$, $(\hat{\alpha}_{2z}, \hat{\alpha}_{1y}, \hat{\alpha}_{3x})$ correspond to the wave functions

$$\psi(y) = \begin{pmatrix} E_z \\ E_x \\ iH_z \\ iH_x \end{pmatrix}, \quad \psi(x) = \begin{pmatrix} E_y \\ E_z \\ iH_y \\ iH_z \end{pmatrix}, \quad \psi(z) = \begin{pmatrix} E_x \\ E_y \\ iH_x \\ iH_y \end{pmatrix} \text{ and to non-zero Poynting vectors}$$

$\psi^+ \hat{\alpha}_{2y} \psi = 2[\vec{E} \times \vec{H}]_y$, $\psi^+ \hat{\alpha}_{2x} \psi = 2[\vec{E} \times \vec{H}]_x$, $\psi^+ \hat{\alpha}_{2z} \psi = 2[\vec{E} \times \vec{H}]_z$ respectively.
As we see, once again we get the correct results.

Now we will prove that the above choice of the matrices and wave functions gives the correct electromagnetic equation forms. Using for example equation (2.16) and transposing the indexes clockwise we obtain for the positive direction of the electromagnetic wave the following results for the $x$, $y$, $z$-directions correspondingly:

$$\begin{cases} \dfrac{1}{c}\dfrac{\partial E_y}{\partial t} + \dfrac{\partial H_z}{\partial x} = -j_y^e \\ \dfrac{1}{c}\dfrac{\partial H_z}{\partial t} + \dfrac{\partial E_y}{\partial x} = j_z^m \\ \dfrac{1}{c}\dfrac{\partial E_z}{\partial t} - \dfrac{\partial H_y}{\partial x} = -j_z^e \\ \dfrac{1}{c}\dfrac{\partial H_y}{\partial t} - \dfrac{\partial E_z}{\partial x} = j_y^m \end{cases}, \quad \begin{cases} \dfrac{1}{c}\dfrac{\partial E_z}{\partial t} + \dfrac{\partial H_x}{\partial x} = -j_z^e \\ \dfrac{1}{c}\dfrac{\partial H_x}{\partial t} + \dfrac{\partial E_z}{\partial x} = j_x^m \\ \dfrac{1}{c}\dfrac{\partial E_x}{\partial t} - \dfrac{\partial H_z}{\partial x} = -j_x^e \\ \dfrac{1}{c}\dfrac{\partial H_z}{\partial t} - \dfrac{\partial E_x}{\partial x} = j_z^m \end{cases}, \quad \begin{cases} \dfrac{1}{c}\dfrac{\partial E_x}{\partial t} + \dfrac{\partial H_y}{\partial x} = -j_x^e \\ \dfrac{1}{c}\dfrac{\partial H_y}{\partial t} + \dfrac{\partial E_x}{\partial x} = j_y^m \\ \dfrac{1}{c}\dfrac{\partial E_y}{\partial t} - \dfrac{\partial H_x}{\partial x} = -j_y^e \\ \dfrac{1}{c}\dfrac{\partial H_x}{\partial t} - \dfrac{\partial E_y}{\partial x} = j_x^m \end{cases}, \quad (5.1)$$

As we can see, we have obtained three equation groups, each of which contains four equations, *as is necessary for the description of all electromagnetic wave directions*. In the same way for all other forms of the Dirac equation analogue results can be obtained.

Obviously, it is also possible via canonical transformations to choose the Dirac matrices in such a way that the electromagnetic wave will have any direction. Let us show it.

## 5.1. The canonical transformations of Dirac's matrices and bispinors

The choice (1.4) of the matrices is not unique [1,3,5]. As it is known, there is a free transformation of a kind: $\alpha = S\, a'S^+$, where $S$ is a unitary matrix, called the canonical transformation operator and also the wave functions $\psi'$ transformation $\psi = S\,\psi'$, which does not change the results of the theory.

If we choose matrices $\alpha'$ as:,

$$\hat{\tilde{\alpha}}_1 = \begin{pmatrix} \hat{\sigma}_x & 0 \\ 0 & \hat{\sigma}_x \end{pmatrix},\ \hat{\tilde{\alpha}}_1 = \begin{pmatrix} \hat{\sigma}_y & 0 \\ 0 & -\hat{\sigma}_y \end{pmatrix},\ \hat{\tilde{\alpha}}_3 = \begin{pmatrix} \hat{\sigma}_z & 0 \\ 0 & \hat{\sigma}_z \end{pmatrix},\ \hat{\tilde{\alpha}}_4 = \begin{pmatrix} 0 & -i\hat{\sigma}_y \\ i\hat{\sigma}_y & 0 \end{pmatrix}, \quad (5.2)$$

then the functions $\psi$ will be connected to functions $\psi'$ according to the relationships:

$$\psi_1 = \frac{\psi'_1 - \psi'_4}{\sqrt{2}},\ \psi_2 = \frac{\psi'_2 + \psi'_3}{\sqrt{2}},\ \psi_3 = \frac{\psi'_1 + \psi'_4}{\sqrt{2}},\ \psi_4 = \frac{\psi'_2 - \psi'_3}{\sqrt{2}}, \quad (5.3)$$

The unitary matrix $S$, which corresponds to this transformation, is equal to:

$$S = \frac{1}{\sqrt{2}} \begin{bmatrix} 1 & 0 & 0 & -1 \\ 0 & 1 & 1 & 0 \\ 1 & 0 & 0 & 1 \\ 0 & 1 & -1 & 0 \end{bmatrix}, \quad (5.4)$$

It is not difficult to check that by means of this transformation we will also receive the equations of the Maxwell theory. Actually, using (2.12) and (5.3) it is easy to receive:

$$\frac{\psi'_1 - \psi'_4}{\sqrt{2}} = E_x,\ \frac{\psi'_2 + \psi'_3}{\sqrt{2}} = E_z,\ \frac{\psi'_1 + \psi'_4}{\sqrt{2}} = iH_x,\ \frac{\psi'_2 - \psi'_3}{\sqrt{2}} = iH_z, \quad (5.5)$$

whence:

$$\psi' = \frac{\sqrt{2}}{2} \begin{pmatrix} E_x + iH_x \\ E_z + iH_z \\ E_z - iH_z \\ -E_x + iH_x \end{pmatrix} \quad (5.6)$$

Substituting these functions in the Dirac equation we will receive the correct Maxwell equations for the electromagnetic waves (in double quantity). It is possible to assume, that the functions $\psi'$ correspond to the electromagnetic wave, moving under the angle of 45 degrees to both coordinate axes.

Thus, from above it follows that *every choice of the Dirac matrices defines only the direction of the initial electromagnetic wave.* Obviously, this is a physical origin why "the physical consequences of Dirac's equation do not depend on the special choice of Dirac's matrices" [4].

## 6.0. The electromagnetic form of the electron theory Lagrangian

As a Lagrangian of the Dirac theory can take the expression [3]:

$$L_D = \psi^+\left(\hat{\varepsilon} + c\hat{\vec{\alpha}}\ \hat{\vec{p}} + \hat{\beta}\ mc^2\right)\psi, \quad (6.1)$$

For the electromagnetic wave moving along the $-y$-axis the equation (6.1) can be written:

$$L_D = \frac{1}{c}\psi^+ \frac{\partial \psi}{\partial t} - \psi^+ \hat{\alpha}_y \frac{\partial \psi}{\partial y} - i\frac{mc}{\hbar}\psi^+ \hat{\beta}\,\psi, \tag{6.2}$$

Transferring each term of (6.2) in the electrodynamics form we obtain the electromagnetic form of the Dirac theory Lagrangian:

$$L_{DM} = \frac{\partial U}{\partial t} + div\,\vec{S} - i\frac{\omega}{4\pi}(\vec{E}^2 - \vec{H}^2), \tag{6.3}$$

(Note that in the case of the variation procedure we must distinguish the complex conjugate field vectors $\vec{E}^*, \vec{H}^*$ and $\vec{E}, \vec{H}$). Using the complex electrical and "magnetic" currents (2.15') and (2.15'') we take:

$$L_{DM} = \frac{\partial U}{\partial t} + div\,\vec{S} - (\vec{j}^{\,e}\vec{E} - \vec{j}^{\,m}\vec{H}), \tag{6.4}$$

It is interesting that since $L_s = 0$ thanks to (1.6), we can take the equation:

$$\frac{\partial U}{\partial t} + div\,\vec{S} - (\vec{j}^{\,e}\vec{E} - \vec{j}^{\,m}\vec{H}) = 0, \tag{6.5}$$

which has the form of the energy-momentum conservation law for the Maxwell equation with current.

## 7.0. About stability of the ring electromagnetic wave. Lorentz's force

This problem is very important for the proof of the nonlinear theory. In classical electron theory there had not been found the forces, which can keep the parts of charge together. Here we can decide on this problem.

In the curvilinear (ring) electromagnetic wave (photon), as result of the interaction between the tangential ring current $\vec{j}_\tau$ and magnetic field of the wave $\vec{H}$, the magnetic component of Lorenz's force density appears, which acts in opposite direction to the Coulomb's electric force:

$$\vec{f}_M = \frac{1}{c}[\vec{j}_\tau \times \vec{H}_p], \tag{7.1}$$

which has the direction along with the ring radius. The Coulomb's force, opposite to said, can be written as:

$$\vec{f}_E = \rho_E \vec{E}_p, \tag{7.2}$$

The full Lorentz force will be:

$$\vec{f} = \rho_e \vec{E}_p - \frac{1}{c}\vec{j}_\tau \times \vec{H}_p = \rho_e E_p \vec{r}^{\,o} - \frac{1}{c} j_\tau H_p \vec{r}^{\,o}, \tag{7.3}$$

where $\vec{r}^{\,o}$ is the unit radius-vector and "p" means «nonlinear electromagnetic wave (photon)». Since $\rho\,c = j_\tau$, we have:

$$\vec{f} = \rho_e (E_p - H_p) \cdot \vec{r}^{\,0}, \tag{7.4}$$

Obviously if the magnetic Lorenz's force is bigger than the Coulomb's force, i.e. the magnetic field is bigger than the electric field

$$H_p > E_p, \tag{7.5}$$

the equilibrium of forces and the stability of twirled photon can appear. From the solution of the Dirac equation [A. G. Kyriakos. http://arXiv.org/abs/quant-ph/0404088] follows that in the normalizing units $H_p = 1$ and $E_p = 1/2$ so that

$$H_p = 2E_p, \tag{7.6}$$

Then the resulting force is

$$\vec{f} = -\rho_e E_p \cdot \vec{r}^0, \tag{7.7}$$

In the case of the electromagnetic wave along the $y$-axis we have two possibilities: $E_p = E_x$ and $E_p = E_z$. According to this fact the force can have two directions: $\vec{f}_x = -\rho_e E_x \cdot \vec{r}^0$ и $\vec{f}_z = -\rho_e E_z \cdot \vec{r}^0$, and $\vec{f}_y = 0$. Let's show how this force in the general theory appears.

The expression of Lorentz's force by the energy-momentum tensor of electromagnetic field $\tau_\mu^\nu$ is well known [6]:

$$f_\mu = -\frac{1}{4\pi}\frac{\partial \tau^\nu_\mu}{\partial x^\nu} \equiv -\frac{1}{4\pi}\partial_\nu \tau_\mu^{\ \nu}, \tag{7.8}$$

This tensor is symmetrical and has the following components:

$$\tau_{ij} = -(E_i E_j + H_i H_j) + \frac{1}{2}\delta_{ij}(\vec{E}^2 + \vec{H}^2),$$

$$\tau_{i0} = 4\pi\, S_p = [\vec{E} \times \vec{H}], \tag{7.9}$$

$$\tau_{00} = 4\pi\, U = \frac{1}{2}(\vec{E}^2 + \vec{H}^2),$$

where $i,j = 1,2,3$, and $\delta_{ij} = 0$ when $i = j$, and $\delta_{ij} = 1$ when $i \neq j$. Using (7.9) it can be written:

$$f_x = f_z = 0, \quad f_y \equiv -\left(\frac{\partial \vec{g}}{\partial t} + \text{grad}\, U\right) \tag{7.20}$$

$$f_0 = -\left(\frac{1}{c}\frac{\partial U}{\partial t} + c\, \text{div}\, \vec{g}\right), \tag{7.21}$$

As we see by using of the symmetrical energy-momentum tensor we don't obtain the needed components of the force.Как мы видим, в случае выбора симметричного тензора энергии импульса мы не получили нужных компонент силы.

The right result can obtain using antisimmetrical spin tensor $\alpha_{\mu\nu}$ (3.1), правильный результат получается, если использовать антисимметричный тензор спина. Then we have:

$$f_\mu = -\frac{1}{4\pi}\frac{\partial \alpha^\nu_\mu}{\partial x^\nu} \equiv -\frac{1}{4\pi}\partial_\nu \alpha_\mu^{\ \nu}, \tag{7.12}$$

or using the tensor components:

$$\begin{cases} f_x = -\left(\dfrac{\partial \alpha_{12}}{\partial x_2} + \dfrac{\partial \alpha_{14}}{\partial x_4}\right) \\ f_y = 0 \\ f_x = -\left(\dfrac{\partial \alpha_{32}}{\partial x_2} + \dfrac{\partial \alpha_{34}}{\partial x_4}\right) \\ f_0 = 0 \end{cases} \quad (7.13)$$

Using (2.12) and (3.1) we obtain of Lorenz's force components:

$$2\pi f_x = E_x\left(\frac{1}{c}\frac{\partial H_z}{\partial t} - \frac{\partial E_x}{\partial y}\right) + H_z\left(\frac{1}{c}\frac{\partial E_x}{\partial t} - \frac{\partial H_z}{\partial y}\right) + H_x\left(\frac{\partial E_z}{\partial t} + \frac{\partial H_x}{\partial y}\right) + E_z\left(\frac{1}{c}\frac{\partial H_x}{\partial t} + \frac{\partial E_z}{\partial y}\right)$$

$$f_y = 0,$$

$$2\pi f_x = E_x\left(\frac{1}{c}\frac{\partial H_x}{\partial t} - \frac{\partial E_z}{\partial y}\right) - H_z\left(\frac{1}{c}\frac{\partial E_z}{\partial t} - \frac{\partial H_x}{\partial y}\right) + H_x\left(\frac{\partial E_x}{\partial t} + \frac{\partial H_z}{\partial y}\right) - E_z\left(\frac{1}{c}\frac{\partial H_z}{\partial t} + \frac{\partial E_x}{\partial y}\right)$$

$$f_0 = 0,$$
(7.14)

For the linear photon all the brackets in (7.14) are equal to zero according to Maxwell's equation. It means that appear no forces in linear photon. When photon rolls up around any of the axis, which are perpendicular to the Y-axis, we will get the additional current terms. Formally we can write

$$\frac{\partial E_x}{\partial t} \to i\omega E_x + \frac{\partial E_x}{\partial t}, \quad (7.15')$$

$$\frac{\partial E_z}{\partial t} \to i\omega E_z + \frac{\partial E_z}{\partial t}., \quad (7.16'')$$

For spinning photon ($E_x$, $H_z$), the force components are (the upper left index shows the spinning axis $OZ$ or $OX$).

$$^z f_x = 2i\frac{1}{4\pi}\frac{\omega}{c}E_x(E_x + H_z) = 2\frac{1}{c}j_\tau \cdot (E_x + H_z) \quad (7.16)$$

for spinning photon ($E_z$, $H_x$):

$$^x f_z = -2i\frac{1}{4\pi}\frac{\omega}{c}E_z(E_z - H_x) = -2\frac{1}{c}j_\tau \cdot (E_z - H_x) \quad (7.17)$$

$$f_y = 0 \quad (7.18)$$

$$f_0 = 0 \quad (7.19)$$

## 8.0. The equation of the electron field motion

We can suppose that 4-vector-potential of electromagnetic field, multiplied to the electron charge $e$, $\left\{e\varphi, \dfrac{e}{c}\vec{A}\right\}$ is the 4-vector of the energy-momentum of the curvilinear wave field $\{\varepsilon_p, \vec{p}_p\}$ (see first part of the paper).

Therefore, the well-known analysis of Dirac's electron equation in the external field can be used for the analysis of the equations of the inner twirled photon field by the change:

$$\frac{e}{c}\vec{A} \to \vec{p}_p, \quad e\varphi \to \varepsilon_p, \quad m \to 0 \tag{8.1}$$

As it is known [1,3], the equation of the electron motion in the external field can be found from the next operator equation, having the Poisson brackets

$$\frac{d\hat{O}}{dt} = \frac{\partial \hat{O}}{\partial t} + \frac{1}{i\hbar}\left(\hat{O}\hat{H} - \hat{H}\hat{O}\right), \tag{8.2}$$

where $\hat{O}$ is the physical value operator, whose variation we want to find and $\hat{H}$ is the Hamilton operator of Dirac's equation.

The Hamilton's operator of the Dirac equation is equal [1,2,3]:

$$\hat{H} = -c\hat{\vec{\alpha}}\,\hat{\vec{P}} - \hat{\beta}\,mc^2 + \varepsilon, \tag{8.3}$$

where $\hat{\vec{P}} = \hat{\vec{p}} - \vec{p}_p$ is full momentum of twirled photon.

For $\hat{O} = \hat{\vec{P}}$ from (8.3) we have:

$$\frac{d\hat{\vec{P}}}{dt} = \left[-\operatorname{grad}(e\varphi) - \frac{e}{c}\frac{\partial \vec{A}}{\partial t}\right] + \frac{e}{c}\left[\vec{v} \times \operatorname{rot} \vec{A}\right], \tag{8.4}$$

or, substitute $\vec{v} = c\hat{\vec{\alpha}}$, where $\vec{v}$ - velocity of the electron matter, we obtain:

$$\frac{d\vec{P}}{dt} = e\vec{E} + \frac{e}{c}\left[\vec{v} \times \vec{H}\right] = f_L, \tag{8.5}$$

Since for the motionless electron $\frac{d\hat{\vec{P}}_p}{dt} = 0$, then the motion equation is:

$$\left(\frac{\partial \vec{p}_p}{\partial t} + \operatorname{grad} \varepsilon_p\right) - \left[\vec{v} \times \operatorname{rot} \vec{p}_p\right] = 0, \tag{8.6}$$

Passing to the energy and momentum densities

$$\vec{g}_p = \frac{1}{\Delta\tau}\vec{p}_p, \quad U_p = \frac{1}{\Delta\tau}\varepsilon_p, \tag{8.7}$$

we obtain the equation of matter motion of twirled photon:

$$\left(\frac{\partial \vec{g}_p}{\partial t} + \operatorname{grad} U_p\right) - \left[\vec{v} \times \operatorname{rot} \vec{g}_p\right] = 0 \tag{8.8}$$

Let us analyse the physical means of (8.8), considering the motion equation of ideal liquid in form of Lamb's-Gromek's equation [12]. In this case, when the external forces are absent, this equation is:

$$\left(\frac{\partial \vec{g}_l}{\partial t} + \operatorname{grad} U_l\right) - \left[\vec{v} \times \operatorname{rot} \vec{g}_l\right] = 0, \tag{8.9}$$

where $U_l, \vec{g}_l$ -energy and momentum density of ideal liquid.

Comparing (8.8) and (8.9) is not difficult to see their mathematical identity. From this follows the interesting conclusion: the microcosm equation may be interpreted as the motion equation of ideal liquid.

According to (8.5, 8.6) from (8.9) we have

$$\frac{\partial \vec{g}_p}{\partial t} + grad\ U_p = \vec{f}_L, \tag{8.10}$$

where $f_L$ is the Lorenz force. As it is known the term $[\vec{\upsilon} \times rot\ \vec{g}_p]$ in (8.9) is responsible for centripetal acceleration. Probably, we have the same in (8.8). If the "photon' liquid" move along the ring of $r_p$ radius, then the angular motion velocity $\omega$ is tied with $rot\ \vec{\upsilon}$ by expression:

$$rot\ \vec{\upsilon} = 2\vec{\omega}_p = 2\omega_p \vec{e}_z, \tag{8.11}$$

and centripetal acceleration is

$$\vec{a}_n = \frac{1}{2}\vec{\upsilon} \times rot\ \vec{\upsilon} = \frac{\upsilon^2}{r_p}\vec{e}_r = c\omega_p \vec{e}_r, \tag{8.12}$$

where $\vec{e}_r$ is unit radius-vector, $\vec{e}_z$ - is unit vector of *OZ*-axis. As a result the equation (5.25) has the form of Newton's law:

$$\rho\ \vec{a}_n = \vec{f}_L, \tag{8.13}$$

This result can be seeing as the electromagnetic representation of the Erenfest theorem [3].

## Conclusion

The above results proof that the nonlinear representation of the Dirac theory give the classical explanation of all particularities of the Dirac electron theory, but they don't contradict to the quantum theory.

(*) Present address: Athens, Greece, e-mail: agkyriak@yahoo.com